\documentclass[12pt, article]{JHEP3}
\usepackage{amsmath,amssymb,euscript,array,mathrsfs,epsfig}
\newcommand{\startappendix}{
\setcounter{section}{0}
\renewcommand{\thesection}{\Alph{section}}}
\newcommand{\Appendix}[1]{
\refstepcounter{section}
\begin{flushleft}
{\large\bf Appendix 
: #1}
\end{flushleft}}
\def\ben
{\begin{equation}}
\def\een{\end{equation}}
\def\be{\begin{equation}}
\def\ee{\end{equation}}
\def\ba{\begin{array}}
\def\ea{\end{array}}

\def\dalemb#1#2{{\vbox{\hrule height .#2pt
        \hbox{\vrule width.#2pt height#1pt \kern#1pt
                \vrule width.#2pt}
        \hrule height.#2pt}}}

\newcommand{\bea}{\begin{eqnarray}}
\newcommand{\eea}{\end{eqnarray}}

\newcommand{\Tr}{{\rm Tr} }

\title{Supersymmetric Lifshitz-like backgrounds from ${\cal N}=4$ SYM with heavy quark density}
\author{Anton F. Faedo, Benjo Fraser and S. Prem Kumar\\\\
{\it 
Department of Physics, \\
Swansea University, \\ 
Singleton Park, Swansea\\
SA2 8PP, U.K.  
}\\
E-mail: \email{ a.f.faedo@swansea.ac.uk, pyfraser@swansea.ac.uk, s.p.kumar@swansea.ac.uk}
} 
\abstract{We examine a class of gravity backgrounds obtained by considering the backreaction of a spatially uniform density of mutually BPS Wilson lines or heavy quarks in ${\cal N}=4$ SUSY Yang-Mills theory. The configurations preserve eight supercharges 
and an $SO(5)$ subgroup of the $SO(6)$ R-symmetry. They are obtained by considering the $\tfrac{1}{4}$-BPS geometries associated to smeared string/D3-brane (F1-D3) intersections. We argue that for the (partially) localized intersection, the geometry exhibits a flow from ${\rm AdS}_5\times {\rm S}^5$ in the UV to a novel IR scaling solution displaying anisotropic Lifshitz-like scaling with dynamical critical exponent $z=7$, hyperscaling violation and a logarithmic running dilaton. We also obtain a two-parameter family of smeared  $\tfrac{1}{4}$-BPS solutions on the Coulomb branch of ${\cal N}=4$ SYM exhibiting Lifshitz scaling and hyperscaling violation. For a certain parametric range these yield IR geometries which are conformal to ${\rm AdS}_2\times {\mathbb R}^3$, and which have been argued to be relevant for fermionic physics.}
\begin{document}
\section{Introduction}
The finite density physics of strongly interacting quantum systems presents significant challenges for existing theoretical frameworks. The questions of physical interest range from the behaviour of quark/baryonic matter at high densities
\cite{Fukushima:2010bq, Rajagopal:2000wf, CasalderreySolana:2011us} to the physics of quantum critical points in condensed matter systems
\cite{Sachdev:2012dq}. The AdS/CFT correspondence \cite{maldacena} provides an avenue for tackling such phenomena within toy models that can be constructed and analyzed using the framework of gauge/string duality
\cite{Hartnoll:2009sz, Herzog:2009xv}.  One of the directions that has  attracted considerable attention in recent years is the possible emergence of 
Lifshitz \cite{Kachru:2008yh, Taylor:2008tg} and Lifshitz-like (with hyperscaling violation)  \cite{Ogawa:2011bz, Huijse:2011ef, Dong:2012se} scaling symmetries as infrared (IR) descriptions of strongly coupled field theories with holographic duals. A particularly fascinating feature of such scaling is the possibility of observing
logarithmic violation of the area law for entanglement entropies indicating the presence of hidden Fermi surfaces \cite{Ogawa:2011bz, Huijse:2011ef}, and other possible signatures of holographic fermionic physics \cite{Hartnoll:2012wm}.

The aim of this paper is to exhibit some supersymmetric backgrounds in string theory resulting from the backreaction of distributions of fundamental strings (F1-strings) in the presence of a large number of D3-branes, giving rise to (IR) geometries exhibiting a range of Lifshitz-like scalings with and without hyperscaling violation\footnote{String/M-theory realizations of backgrounds with hyperscaling violation (e.g. from lightlike compactification of AdS plane waves) and their properties have been studied in \cite{Narayan:2012hk, Singh:2012un}.}. Our primary motivation for considering such configurations arises from the interpretation of macroscopic fundamental strings as heavy quarks or Wilson lines \cite{Maldacena:1998im} in the holographic dictionary between strongly coupled  ${\cal N}=4$ SUSY Yang-Mills theory in the 't Hooft large-$N$ limit, and IIB string theory in ${\rm AdS}_5 \times
 {\rm S}^5$. Spatially uniform distributions of macroscopic strings correspond to a finite density of heavy quarks in the gauge theory\footnote{It can also be argued that such configurations in the bulk should control the IR geometry produced by the backreaction of massive and massless flavour D7-branes with chemical potential for baryon number. We thank D. Mateos and J. Tarrio for stimulating discussions on this point.}. It has been found in \cite{Kumar:2012ui} that a specific non-supersymmetric smearing of such string sources, preserving global $SO(6)$ symmetry, produces a flow from ${\rm AdS}_5 \times
 {\rm S}^5$ to a Lifshitz geometry with critical exponent $z=7$ and a logarithmically running dilaton (see also \cite{Azeyanagi:2009pr}). 

Spatially homogeneous $(d+2)$-dimensional spacetime metrics exhibiting Lifshitz-like scaling with dynamical critical exponent $z$ and hyperscaling violation have the form \cite{Ogawa:2011bz, Huijse:2011ef, Dong:2012se}
\be
ds^2\,=\,r^{-2\theta/d}\,\left(-r^{2 z}\,dt^2 \,+\,\frac{dr^2}{r^2}\,+\,r^2{d\vec x^{\,2}}\right), 
\ee
so that under the rescalings: $\vec x\mapsto a\, \vec x$, $r\mapsto a^{-1}\,r$ and $t\mapsto a^z\, t$, the line element transforms covariantly as $ds\mapsto a^{\theta/d}\,ds$. 

We will encounter such scaling backgrounds with $d=3$, preserving 8 supersymmetries, as IR descriptions of 
${\cal N}=4$ SYM at strong coupling with a smeared density of supersymmetric Wilson lines or heavy quarks. In the large-$N$ limit, we take the number density of such heavy quarks to scale as $N^2$, so that it becomes necessary to include their backreaction on the field theory vacuum. The holographic duals to such states are described by a class of $\tfrac{1}{4}$-BPS backgrounds in IIB supergravity whose general properties will be presented elsewhere \cite{paper2}. The specific class of backgrounds that we consider in this article are those which have the form of certain well known intersecting brane configurations \cite{Tseytlin:1996bh,Gauntlett:1997cv,Youm:1999zs}. The configurations within this class can be thought of as pairs of quarks and antiquarks which are mutually supersymmetric (by having opposite internal $SO(6)$ orientations) and are distributed uniformly in the spatial directions of the gauge theory. 
A particular solution within  this class has appeared in 
\cite{Singh:2012un, deyroy} where its Lifshitz-like scaling properties have been pointed out.

We find that one category of the intersecting F1-D3 configurations
naturally leads to an exact scaling solution in the infrared, exhibiting 
Lifshitz-scaling with $z=7$ which is mildly broken by a logarithmically running dilaton. The background preserves supersymmetry and an $SO(5)$ internal symmetry. This result is noteworthy since the $z=7$ Lifshitz scaling behaviour matches   that of \cite{Kumar:2012ui} which dealt with an $SO(6)$ preserving non-supersymmetric setup. This is indicative that the $z=7$ scaling is probably a universal feature of  ${\cal N}=4$ SYM with finite quark density (at strong coupling and large-$N$). It points toward the possibility that when the quark flavours are made dynamical (by introducing D7-branes for instance \cite{karchkatz}) and their backreaction taken into account at finite baryon density, then the low temperature IR physics  may well be controlled by a similar scaling solution (see \cite{Hartnoll:2009ns} for closely related discussions). A concrete framework where this possibility can be further investigated is the unquenched smeared flavour brane setup of \cite{Benini:2006hh, Bigazzi:2011it,Bigazzi:2013jqa, CarlosReview}.

Within the setting of the supersymmetric configurations in this paper, constructing the full flow from ${\rm AdS}_5\times {\rm S}^5$ to the IR scaling solution remains a challenging task due to the reduced isometry of the dual gravity backgrounds. However, we analytically examine the departure of the flow induced by the supersymmetric smeared strings in the UV, and find close similarities to the $SO(6)$ symmetric case of \cite{Kumar:2012ui} wherein it was possible to obtain the entire flow numerically.

The supersymmetric intersecting brane configurations present us with a further interesting route for obtaining non-trivial scaling solutions. This is because of the existence of supersymmetric moduli spaces at zero temperature. In particular, it is possible to move out to a generic point on the Coulomb branch moduli space (where the gauge group is generically Higgsed to a product of Abelian factors) of ${\cal N}=4$ SYM and examine the effect of a finite quark density. In the dual gravity description this corresponds to a general distribution of $N$ D3-branes at large-$N$. We consider such continuous distributions of the D3-branes and F1-strings that are compatible with the supersymmetries preserved by the intersecting brane configurations. Specifically, we introduce distributions for these sources with power law scalings in the IR and solve the supergravity equations for such power law density functions. We thus find a two parameter family of Lifshitz-like scaling solutions with hyperscaling violation coefficients. Imposing the weak energy condition on each of the source distributions restricts the allowed values of $z$ and $\theta$. 

Perhaps the most interesting result within the class of Coulomb branch configurations is the appearance of a solution with $z,\theta\to\infty$ and $\eta\,\equiv\,-\theta/z$ fixed (satisfying $\eta\geq 1$). These give rise to geometries that are conformal to ${\rm AdS}_2\times {\mathbb R}^3$ with vanishing entropy at zero temperature. Such geometries have been argued to be relevant for holographic descriptions of fermionic physics \cite{Hartnoll:2012wm}.

All the solutions we find bear close resemblance to scaling solutions in Einstein-Maxwell-dilaton theories \cite{Taylor:2008tg, Goldstein:2009cv,
Goldstein:2010aw, Charmousis:2010zz}. In particular the dilaton softly breaks the scaling invariance via logarithmic running. Furthermore since the dilaton runs to zero in the IR, it  
renders $\alpha^\prime$ corrections important in the deep IR.

\section{Smeared strings and ${\cal N }=4$ SYM} 

${\cal N}=4$ SYM theory possesses a global $SO(6)$ R-symmetry which corresponds to the isometry of the five-sphere of the dual gravity background at strong coupling. A supersymmetric Wilson line in ${\cal N}=4$ SYM at strong coupling is computed by an infinite probe F1-string stretching from the boundary of ${\rm AdS}_5$ along the radial AdS coordinate, whilst remaining localized at a point on the internal $S^5$
\cite{Maldacena:1998im}. Such a configuration preserves $SO(5)\subset SO(6)$ and more generally, is associated to the Maldacena-Wilson line in some representation ${\cal R}$,
\be
W_{\cal R}(\vec x)\,=\,\frac{1}{{\rm dim[{\cal R}]}}{\Tr}_{\cal R}\,{\cal P}\exp\left(i\int dt\,(A_0 (\vec x, t)\,+\, \hat n^I\,\phi_I(\vec x, t))\right)\,,
\ee
where $\hat n$ is a unit vector in ${\mathbb R}^6$ and $\phi_I$, the six adjoint scalars in the ${\cal N}=4$ multiplet. We focus our attention on straight (timelike) BPS-Wilson lines. There also exists a gravity dual realization of such supersymmetric or BPS-Wilson lines for generic representations ${\cal R}$ in terms of D3 and D5-brane embeddings in ${\rm AdS}_5 \times
 {\rm S}^5$ 
\cite{Drukker:2005kx, yamaguchi, Gomis:2006sb, Hartnoll:2006is}. In particular, for representations whose size scales 
as $N^2$, the gravity duals of the Wilson lines necessarily require inclusion of back-reaction of the corresponding D3/D5-brane configurations on ${\rm AdS}_5 \times
 {\rm S}^5$, leading to the $\tfrac{1}{2}$-BPS `bubbling' geometries of \cite{D'Hoker:2007fq}.

Our goal in this paper is to model a particular state with finite heavy quark density by taking a very specific BPS Wilson line configuration and smearing it uniformly along all spatial directions of the ${\cal N}=4$ theory. The unsmeared configuration consists of a mutually BPS quark-antiquark pair i.e. antiparallel lines placed at antipodal points of the internal $S^5$ \cite{Maldacena:1998im, Drukker:2011za}. This preserves an $SO(5)$ global symmetry and, prior to taking the near horizon limit of the D3-branes, can be viewed as a  supersymmetric 
F1-D3 intersection. A specific delocalized version of this intersection has been discussed in \cite{Singh:2012un, deyroy} and shown to lead to Lifshitz-like scaling. The backreacted geometries resulting from the spatially smeared configurations  are $\tfrac{1}{4}$-BPS, preserving 8 real supercharges.

We will consider two distinct classes of $\tfrac{1}{4}$-BPS configurations in this paper: 
\begin{itemize}
\item{
In the first category lies the so-called {\em partially localized} F1-D3 intersection \cite{Youm:1999zs}, where F1-strings are smeared along the relative transverse directions (spatial coordinates) on the D3-branes, and the D3-branes are not smeared. We will argue that the corresponding gravity background represents a flow from ${\rm AdS}_5\times{\rm S}^5$ to a $z=7$ Lifshitz-like geometry.}

\item{The second class involves delocalized D3-brane and F1-string distributions, allowing for a general smearing density for both sets of sources. The solution of \cite{Singh:2012un,deyroy} is a special case within this category of solutions. We find a large class of zero temperature scaling solutions with a range of dynamical critical exponents and hyperscaling violation coefficients.}
\end{itemize}

\subsection{Metric ansatz in 10D}

Given the symmetries and supersymmetries of such configurations,  we show in an accompanying paper \cite{paper2}, that the type IIB fermionic variations vanish if one takes the following ansatz for the Einstein frame metric, the Ramond-Ramond fluxes $F_3, F_5$ and Neveu-Schwarz field strength $H_3$:  
\begin{eqnarray}
ds^2_{\rm Einstein}&=&-e^{2\left(A+\phi\right)}dt^2+e^{2A}\sum_{i=1}^3dx^idx^i+e^{-2A+\phi}dx^2+e^{-2A-\phi}\left(dy^2+y^2d\Omega_4^2\right)\,,\nonumber\\[1.3mm]
g_s F_5&=&\frac{y^4}{4}\left(1+*\right)\left[\partial_y\left(e^{-4A-\phi}\right)dx-\partial_x\left(e^{-4A-3\phi}\right)dy\right]\wedge \epsilon_{(4)}\,,\nonumber\\[4.5mm]
H_3&=&\partial_y\left(e^{2\phi}\right)\,dt\wedge dy\wedge dx\,,\quad\quad\quad F_3\,=\,2\,e^{4A-\phi}\partial_x\phi\,dx^1\wedge dx^2\wedge dx^3\,.\label{ansatz}
\end{eqnarray}
Note that we follow the conventions of \cite{D'Hoker:2007fq} and our normalization of $F_5$ differs from the usual one by a factor of 4.
Here $d\Omega_4^2$ is the metric of the unit radius four-sphere with volume form $\epsilon_{(4)}$, the warp factor $A$ and dilaton $\phi$ are functions of the $x$ and $y$ coordinates only.
The gauge theory lives in the 4D spacetime spanned by the coordinates $(t,x^1,x^2, x^3)$. 
In addition, the ten-dimensional complex spinor $\epsilon$ parametrising supersymmetry variations must satisfy the projection
conditions  
\begin{equation}
\Gamma^{\,\hat t\, \hat x }\,D^{-1}\,\epsilon^*\,=\,\epsilon\,,\qquad\qquad
i\,\Gamma^{\,\hat t\, \hat x_1 \hat x_2 \hat x_3 }\epsilon\,=\,\epsilon\,,
\end{equation}
leading to $\tfrac{1}{4}$-BPS solutions. Here $D$ is the ten dimensional complex conjugation matrix. The projections are those associated to fundamental strings and D3-branes, respectively.

Within this ansatz, the geometry can be viewed locally as a fibration of ${\mathbb R}_t\times {\mathbb R}^3\times S^4$ over the $x$-$y$ plane. Therefore, these are time independent backgrounds with an $SO(5)$ isometry and symmetry under translations and rotations in ${\mathbb R}^3$.

\subsection{(Partially) Localized F1-D3 intersection}

In this paper we will only study solutions with $F_3 = 0$. This precludes a charge density for D5-branes that could be interpreted as baryon density as in \cite{Kumar:2012ui}.
It is also physically clear why this is the case: we are considering a uniform density of quark and {\em anti}-quark pairs (with opposite $SO(6)$ orientations).
 With this choice, the backgrounds represent F1-D3 BPS intersections  of the kind analyzed in \cite{Tseytlin:1996bh, Gauntlett:1997cv, Youm:1999zs}. We can make contact with the standard form of the metric for such backgrounds with the identifications:
\be
e^{-2\phi}\,=\,h_1\,,\qquad\qquad e^{-4 A -\phi}\,=\, h_3\,,
\ee
where $h_1$ and $h_3$ are the harmonic functions associated to the F1-strings and D3-branes respectively. It is well known that in these
 types of
solutions (e.g. \cite{Youm:1999zs}) there is a smearing along  the relative transverse directions of the strings/branes i.e. those which are parallel to the world volume of some of the branes (strings), but perpendicular to others. They are constructed by the so-called Ôharmonic ruleÕ: to each type of brane (string) there is a corresponding function which is harmonic in the space transverse to their world-volumes. In order to source two sets of branes in the same solution one simply multiplies the two harmonic functions together in the appropriate way. 

\paragraph{\underline{ ${\bf{\rm\bf AdS}_5 \times
 {\rm\bf S}^5}$  vacuum:}} The vacuum solution, namely ${\rm AdS}_5\times {\rm S}^5$, is recovered when
\be
h_1\,=\,1\,,\qquad  h_3\,=\,e^{-4A}\,=\,\frac{1}{(x^2+y^2)^2}\,, \quad x\,=\, r\cos\theta\,,\quad y\,=\, r\sin\theta\,,\quad 0\leq \theta \leq \pi\,.
\ee
This is indicated in Fig.1. The five-sphere is obtained by fibering the $S^4$ along a semicircle enclosing the origin. 
\begin{figure}[h]
\begin{center}
\includegraphics[width=2.3in]{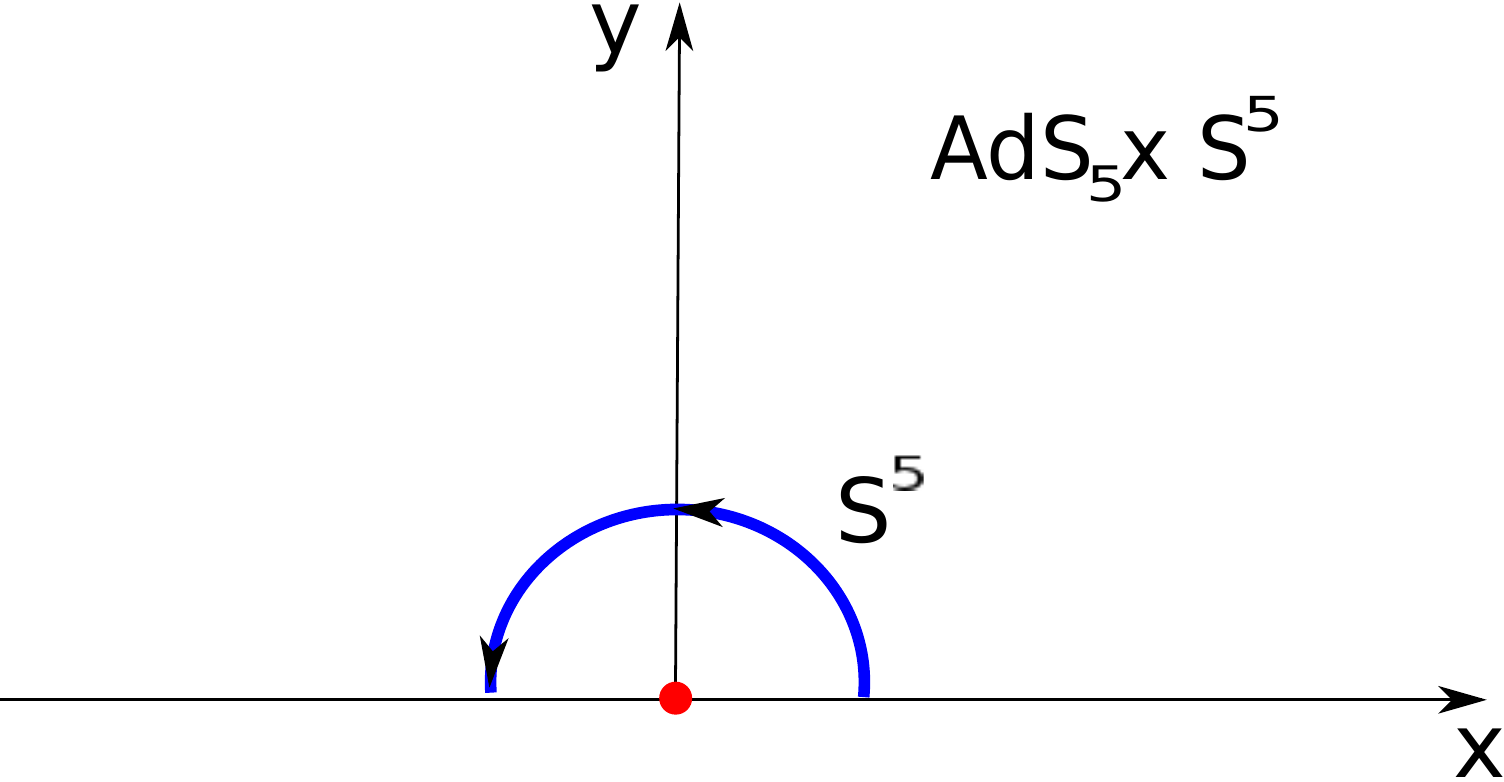}\hspace{1.2in}
\includegraphics[width=2.3in]{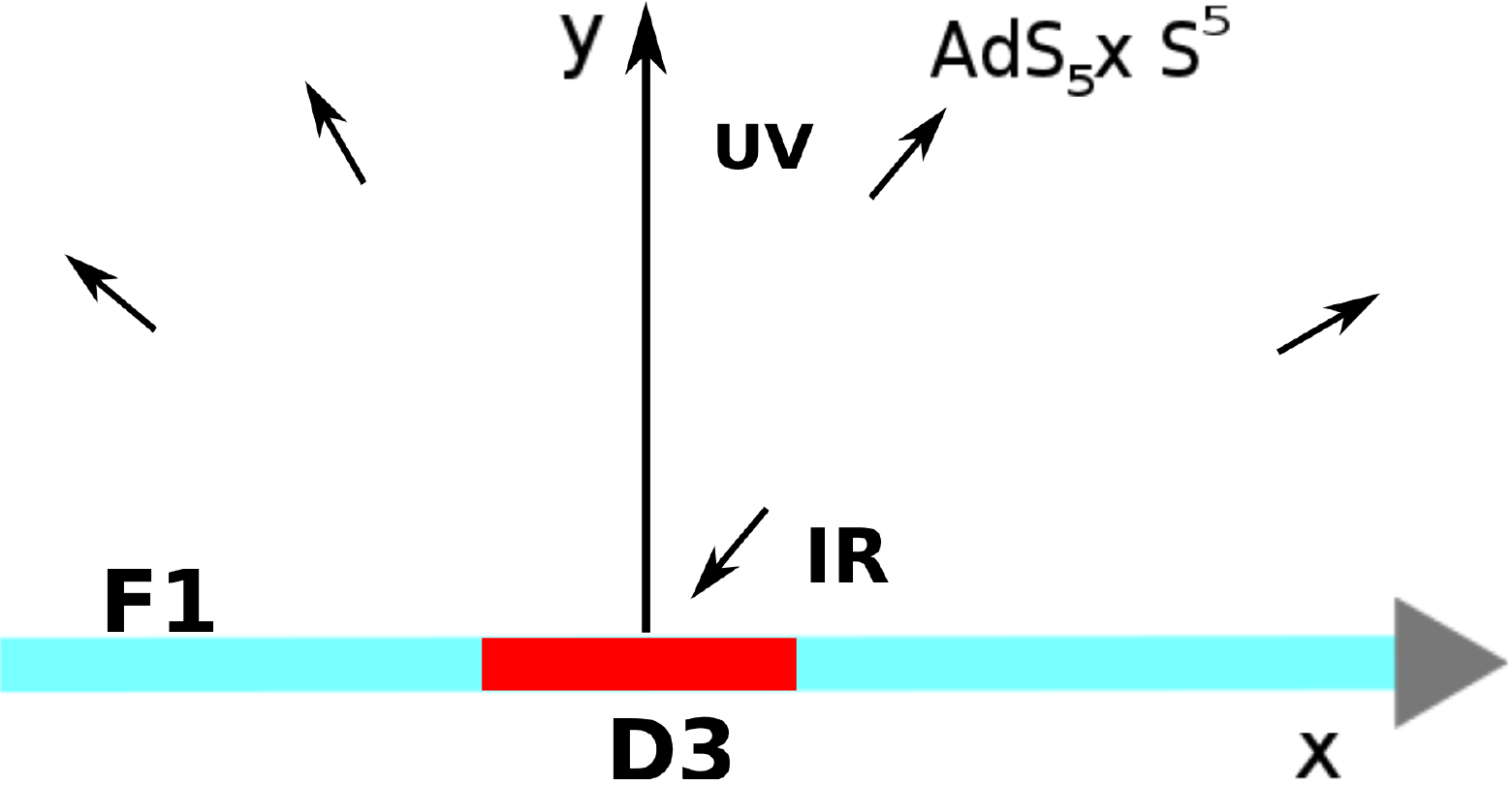}
\end{center}
\caption{\small{{\underline {\bf Left:}} The $x$-$y$ half-plane in AdS
with $x=r\cos\theta$ and 
$y= r\sin\theta$ where $r$ is the radial coordinate in AdS and $\theta$ the polar angle on the $S^5$. {\underline {\bf Right:}} D3-brane positions (in red) on the $x$-$y$ plane, with infinite F1-strings oriented along the $x$-axis. The thickened lines indicate that the respective distributions may have a non-zero extent along the $x$-axis.
}}
\label{xyplane} 
\end{figure}

\paragraph{\underline{Backreaction of string sources:}}
The presence of macroscopic string sources will lead generically to a non-vanishing NS-NS three-form flux $H_3$. The form of $H_3\,=\,dB_2$ in our ansatz \eqref{ansatz} suggests  that the strings must lie parallel to the $x$-axis (see Fig.\ref{xyplane}).
Now, both the functions $h_3$ and $h_1$ are non-trivial, and for 
sources localized in the $x$-$y$ plane they are determined by the Bianchi identity for $F_5$ and equation of motion for $H_3$, namely,
\be
\partial_y\left(y^4\partial_y h_3\right)\, +\,y^4\,h_1 \partial_x^2 h_3\,=\,0\,,\qquad
\qquad \partial_y(y^4 \partial_y h_1)\,=0\,.
\label{h3}
\ee 
These are Laplace-like equations for (partially) localized sources smeared along certain directions transverse to them (the transverse $S^4$ and the spatial ${\mathbb R}^3$). 
Since $F_3=0$, we also know that $h_1$ is only a function of the $y$-coordinate,
\be
e^{-2\phi}\,=\,h_1(y)\,=\,1\,+\,\frac{Q_1}{y^3}\,.
\label{h1}
\ee
This is what we would expect for a localized, infinite string source oriented along the $x$-axis. The additive integration constant has been set to unity by the requirement 
that the dilaton $\phi$ vanishes for large $y$. Here $Q_1$ is proportional to the number density of F-strings $n_{\rm F1}$, which also measures the density of heavy quark sources in the gauge theory:
\be
Q_1\,\equiv\,\sqrt\lambda\,\,\frac{n_{\rm F1}}{N^2}\,\frac{4\pi^2}{{\rm Vol}(S^4)}\,,
\ee
$\lambda$ being the 't Hooft coupling of the ${\cal N}=4$ theory. The number density of heavy quarks $n_{\rm F1}$ must scale as $N^2$ in order to keep $Q_1$ fixed in the large-$N$ limit, and to consistently include their backreaction on the background.

The normalizations are fixed by demanding that the equation of motion for $H_3$ yield a number density  $n_{\rm F1}$ of macroscopic F-strings localized at $y=0$,
\be
d(e^{-\phi} *H_3)\,=\,\frac{n_{\rm F1}}{2\pi\alpha^\prime}\,16\pi G_N\,\delta(y)\,dy\wedge\epsilon_{(4)}\wedge dx^1\wedge dx^2\wedge dx^3\,.
\ee
Setting the AdS-radius to unity, the 't Hooft coupling $\lambda$ of the gauge theory is related to the tension of a fundamental string as $(2\pi\alpha^\prime)^{-1} = \frac{\sqrt\lambda}{2\pi}$. In addition,  Newton's constant in ten dimensions is given by $16\pi G_N\,=\,(2\pi)^7 g_s^2\,\alpha^{\prime\, 4}$, with $\lambda \,=\, 4\pi g_s N$.

 With this solution for $h_1$, the full background is determined by $h_3(x,y)$, which solves eq.\eqref{h3} subject to the requirement that the geometry is asymptotically, locally ${\rm AdS}_5\times{\rm S}^5$. It is not possible to solve this equation analytically, but we can analyze its UV and IR asymptotics.

\subsection{$z=7$ Lifshitz IR}
 In the IR limit, which corresponds to small $y$, we have that $h_1\approx Q_1/y^3$. The general solution for the harmonic function $h_3(x,y)$  in this limit is (e.g. \cite{Youm:1999zs}): 
\bea
&& h_3(x,y)\,=\,\frac{1}{y^{5/2}}\left(c_1\left(v\,+\, 4 Q_1\right)^{5/2}\,+\, c_2\,(v^{5/2}\,+\,10 Q_1\,v^{3/2} \,+\, 30 Q_1^2 \sqrt{v})\,\right)\,,\nonumber\\
&& v\,\equiv\,x^2y\,.
\eea
We may view this as a `near horizon' limit for the F-strings.
In fact $h_3$ can take the more general form, similar to that for  multi-centred D3-brane sources, $h_3\sim \sum_i P_i \left((x-x_i)^2 + {4 Q_1}/{y}\right)^{5/2}+\ldots$, but this generalization is not of particular interest to us at this point. It is easy to see that the solution for $h_3$ follows from a scaling 
argument\footnote{Allowing for a source density $\rho(x,y)$ on the right hand side of \eqref{h3}, we expect that under $x\mapsto s x$, $y\mapsto s^a y$, for a scaling solution we should have $\rho\mapsto\rho/s^{1+a}$. Comparing the two sides we find that $a=-1/2$ and $h_3\,=\,\tilde h(x^2 y)/y^{5/2}$ for some function $\tilde h$. The differential equation for $\tilde h(v\equiv x^2y)$, yields the general solution $\tilde h(v)\,=\,
c_1\,(v+4 Q_1)^{5/2}\,+\,c_2\,\sqrt{v}(v^2+10 Q_1 v + 30 Q_1^2)$.}.
The second important remark is that such solutions cannot correspond to localized F1-D3 intersections, since $h_3$ increases as a positive power of $x$ for large $x$. The D3-branes are in fact delocalized along the $x$-axis in this limit ($y/{Q_1}^{1/3} \ll 1$). The reasons behind such `spontaneous delocalization' were explored in \cite{Marolf:1999uq}, including the case which is precisely the S-dual of the setup we are studying, namely D1-branes smeared along the spatial directions of D3-branes. In this case the spontaneous delocalization was attributed to the analogue of the Coleman-Mermin-Wagner theorem in the 0+1 dimensional quantum mechanics of the intersection.

One can confirm that the D3-branes are indeed not localized by noting that $F_5$ (in this limit and with $c_2=0$ for simplicity) depends only on $v\equiv x^2y$, 
\bea
&&g_sF_5\,=\,(1+*)\,f(v)\,dv\wedge\epsilon_{(4)}\,,\qquad\qquad v\equiv x^2 y\,,\\\nonumber\\\nonumber
&& f(v)\,\equiv\,-\frac{5Q_1}{4\sqrt v}\,(4Q_1+ v)^{5/2}\,c_1\,.
\eea
Therefore, the D3-brane charge can be computed by  $\int_\gamma f(v)dv$ where $\gamma$ is a path joining two points on the curves, say, $v=v_1$   and $v=v_2$.  For any two points on the $x$-axis, which sits at $v=0$, the integral along a path joining them vanishes trivially, and therefore the D3-brane charge is not localized at the origin, nor at any other finite point on the axis. It also follows, by allowing arbitary deformations of the curve, that there are no localized D3-brane sources for any finite values of $x$ and $y$. In fact the sources must be  viewed as being placed at infinity. This is also related to the fact that the $S^4$ in the geometry does not shrink anywhere in the $x$-$y$ plane in this IR limit.
This should be contrasted with ${\rm AdS}_5\times {\rm S}^5$ wherein the D3-brane charge is obtained by choosing a path with end-points on the $x$-axis and, importantly, enclosing the origin, as in Fig.\ref{xyplane}.

This picture therefore suggests that the F-strings have pulled apart the D3-branes so that for scales set by $y\ll {Q_1}^{1/3}$, they appear to be `localized' at infinity. Within the full flow from the UV AdS asymptotics we  expect the D3-branes  to have a non-trivial distribution (and to therefore be delocalized) along the $x$-axis. 
Since the bulk geometry depends on two holographic directions, $y$ and $x$ (or $y$ and $v$),
the D3-brane distribution will emerge from numerical analysis of the entire flow, a study which we postpone for the future.

Plugging in the expressions for $h_3$ and $h_1$ in this IR limit,
defining a new radial coordinate $\rho\equiv y^{1/4}$ and after appropriate coordinate rescalings, we find
\bea
ds^2_{\rm Einstein}\,&=&\, Q_1^{3/2}\,\left[\left( -\,\rho^{14} \,dt^2 
\,+\, \rho^2\, dx^idx^i\right)\, \tilde f(x^2\rho^4)^{-1/2}
\,+\,16\,\frac{d\rho^2}{\rho^2}\,\tilde f(x^2\rho^4)^{1/2}\right.\nonumber\\\nonumber\\
&&\,+\,(d\Omega_4^2\,+ \rho^4 dx^2)\,\tilde f(x^2\rho^4)^{1/2}\left.\right]\,,\label{irlimit1}\\\nonumber\\
e^\phi\,&=&\,\rho^6\,,\nonumber\\\nonumber\\\nonumber
\tilde f(x^2\rho^4)\,&\equiv&\, c_1(x^2\rho^4+4)^{5/2}\,+\,c_2\,((x^2\rho^4)^{5/2}\,+\,10 \,(x^2\rho^4)^{3/2} \,+\, 30 \sqrt{x^2\rho^4})
\eea
This is an exact, supersymmetric solution to the type IIB equations. The constants $c_1$ and $c_2$ can only be determined upon embedding this IR solution into the complete flow, with AdS UV asymptotics, triggered by the string charge density. 
The IR background metric exhibits a scaling symmetry under
\be
x^i\mapsto a\,x^i\,,\qquad t\mapsto a^7\, t\,,\qquad \rho \mapsto a^{-1}\, \rho\,,\qquad x\mapsto a^2\, x\,,
\ee
which is an anisotropic Lifshitz-like scaling, with dynamical critical exponent $z=7$, in the 4D gauge theory. Note that the scaling symmetry is not exact, as it is broken by the logarithmic running of the dilaton with energy scale. 

Let us point out certain noteworthy aspects of the scaling solution \eqref{irlimit1}. The IR Lifshitz scaling is realized in an unusual fashion due to the non-trivial dependence of the metric on two holographic directions, $x$ and $\rho$, and this will further affect the scaling properties of observables such as the entanglement entropy and  thermodynamic quantities. It is remarkable that the $z=7$ Lifshitz scaling, with exactly the same running of the dilaton, was also observed in \cite{Kumar:2012ui, Azeyanagi:2009pr} within the non-supersymmetric $SO(6)$-symmetric ansatz for smeared Wilson lines in ${\cal N}=4$ SYM. This suggests that the emergent Lifshitz symmetry with $z=7$ may be a universal feature of a dense state of heavy quarks introduced in ${\cal N}=4$ SYM, independent of the details of their internal $SO(6)$ orientation. A further difference between the $SO(6)$-symmetric setup of \cite{Kumar:2012ui, Azeyanagi:2009pr} and the supersymmetric configuration above, is that whilst the former has $F_3\neq 0$ corresponding to D5-brane/baryon charge density, the latter has $F_3 =0$. However this detail does not appear to affect the dynamical critical exponent in the  IR. A supersymmetric configuration with $F_3\neq 0$ is explored in 
\cite{paper2}.

We can recover a more conventional form of the Lifshitz-like metric by taking a further limit of the IR solution, namely $x^2\rho^4 \ll 1$, assuming that $c_1\neq 0$ generally. After some coordinate rescalings, this yields
\bea
ds^2_{\rm Einstein}\,&\approx&\, c\left(\frac{d\rho^2}{\rho^2}\,-\,\rho^{14}dt^2\,
+\,\rho^2 \sum_{i=1}^{3}dx^idx^i\, +\, \rho^4 dx^2\,+\,\tfrac{1}{16}d\Omega_4^2\right)\nonumber\\\nonumber\\
 e^{\phi}\,&=&\,\rho^6\,,\label{irlimit}
\eea
where $c$ is a constant. 
The four-sphere has a constant size in this limit, and the background \eqref{irlimit} is not, by itself, an exact solution to the supergravity equations of motion; it is a limiting form of the IR (small $y$) metric. 

An interesting aspect of the IR physics that is manifest in this limit,  is that it is effectively a six-dimensional geometry (after reduction on $S^4$).
The emergence of an extra non-compact coordinate in the IR geometry suggests that the IR field theory of the smeared heavy quark impurities in ${\cal N}=4$ SYM has an emergent dimension. Such an interpretation would be consistent with the presence of a delocalized distribution of D-branes that deconstruct an extra dimension \cite{ArkaniHamed:2001ie}. Given the complicated nature of the putative RG flow solution, and the dual field theory being at strong coupling, it is not easy to pin-point a direct test of this proposal. Our study of smeared strings on the Coulomb branch in the next section, lends support to  this picture. It is known that Coulomb branch configurations in ${\cal N}=4$ SYM deconstruct a higher dimensional field theory \cite{HoyosBadajoz:2010td}. Within such an interpretation the deconstructed dimension should be viewed as being compact, its size being determined by the inverse spacing between two adjacent D-branes. The anisotropy between the deconstructed coordinate $x$ and the spatial coordinates $\vec x$ is automatic due to the distribution of heavy quark sources.

\paragraph{\underline{Entanglement entropy}:} An important probe of the IR physics of the gauge theory is the entanglement entropy, which can be computed using the Ryu-Takayanagi prescription \cite{Ryu:2006bv}. For the metric in eq.\eqref{irlimit1}, this is not a straightforward calculation since the components depend both on $x$ and $\rho$ coordinates and the solution for the extremal surface (with a prescribed boundary)  could depend non-trivially on both these dimensions.

We can, however, make a simplifying approximation by assuming that the physics in the deep infrared should be governed by sufficiently small $\rho$ and $x$, so that $x^2\rho^4 \ll 1$. Then we can use the approximate form \eqref{irlimit} to calculate the entanglement entropy of a `strip' in ${\mathbb R^3}$, specified by
\be
-\ell\leq x^1 \leq \ell\,,\qquad 0\leq x^{2,3} \leq L\,, 
\qquad 0 \leq x\leq \tilde L\,,
\ee 
with $\ell \ll L, \tilde L$ at a UV-slice of the geometry, $\rho =\rho_\Lambda$.
The effective area functional for the 3D surface $\Sigma_3$ with the strip as its boundary, can be defined as \cite{Ryu:2006bv}
\be
{\cal S}\,=\,\frac{1}{4G_N}\int_{\Sigma_3\times S^4\times
{\mathbb R}_x}d^8x\,\sqrt{{\rm det}\,{}^* g}\,.
\ee
Extremizing the functional and extracting its finite part in the usual way (e.g. \cite{Ogawa:2011bz,Dong:2012se})
we find
\be
{\cal S}\Big|_{\rm finite}\,\sim\, N^2 \tilde L \left(\frac{L}{\ell}\right)^2\ell^{-2}\,,
\ee
where we have omitted various constants of proportionality and traded Newton's constant in ten dimensions $G_N$ for a factor of  $N^2$ according to the holographic dictionary. The scaling of the entanglement entropy with $\ell$ is characteristic  of hyperscaling violation with negative $\theta$ \cite{Ogawa:2011bz, Dong:2012se}. If we view the (IR) gauge theory as having 3 spatial dimensions (e.g. with the $x$-coordinate compactified), then $\theta=-2$.

The hyperscaling violation could also be directly inferred by reducing the ten dimensional metric \eqref{irlimit} to five dimensions, by assuming the $x$-coordinate to be compact. The reduction to five dimensions yields
\be
ds^2_5\,=\,\rho^{4/3}\,\left(\frac{d\rho^2}{\rho^2}\,-\,\rho^{14}dt^2\,+\,\rho^2d\vec x^2\right)\,,
\ee
which exhibits Lifshitz scaling with $z=7$ and hyperscaling violation with $\theta=-2$. For the configurations on the Coulomb branch studied below, the emergence of the extra spatial coordinate $x$  can be interpreted via deconstruction and the size of this dimension is controlled by the inverse spacing between D3-branes
\cite{HoyosBadajoz:2010td,ArkaniHamed:2001ie}. 

We stress that the calculation of the entanglement entropy outlined above is only an approximation. It would be interesting to have a more precise computation in the exact scaling background \eqref{irlimit1}, and to check whether these qualitative expectations  are reproduced.

\subsection{UV AdS asymptotics}

We will now attempt to understand how the solution to eq.\eqref{h3} will modify the UV  AdS asymptotics. In order to find a solution which asymptotes to ${\rm AdS}_5\times {\rm S}^5$, and also includes string sources,  we need to use the general form \eqref{h1} for $h_1(y)$. Whilst the problem is not analytically tractable, we can make progress by noting that eq.\eqref{h3} exhibits translational invariance in $x$. Fourier transforming with respect to this variable \cite{Arapoglu:2003ah} yields
\be
\tilde h_3^{\prime\prime}(y;p)\,+\,\frac{4}{y}\,\tilde h_3^{\prime}(y;p)\,-\,p^2\,h_1(y)\,\tilde h_3(y;p)\,=\,0\,,\qquad\qquad h_1(y)\,=\,1+\frac{Q_1}{y^3}\,,
\ee
where primes denote derivatives with respect to $y$.
The equation has irregular singular points of order $1/2$ at $y=0$ and order $1$ at $y=\infty$. Around either of these points, 
the equation can be solved as a power series using a so-called Thom\'e expansion \cite{slavlay}. Around $y=\infty$, this leads to solutions of the type,
\be
\tilde h_3(y;p)\,=\,\frac{e^{py}}{y^2}\,\sum_{n=0}^\infty\frac{a_n(p)}{y^n}\,,
\ee
where the coefficients can be determined via a recursion relation and $p$ can be either positive or negative. A slightly different expansion which has overlap with the large $y$ limit is a formal series expansion in powers of $Q_1$ -- the string or  `heavy quark' density:
\be
\tilde h_3(y;p)\,=\,\sum_{n=0}^\infty Q_1^n \,f_n(y, p)\,.
\ee
Substituting into \eqref{h3}, and solving the resulting equations order by order in $Q_1$, we find,
\bea
&&f_0(y,p)\,=\, \frac{e^{py}}{y^2}\,\left(1 -\frac{1}{py}
\right)\,,\qquad f_1(y,p)\,=\, - \frac{p\,e^{ py}}{4\,y^4}\,,
\\\nonumber\\\nonumber
 &&f_2(y,p)\,=\, -\frac{p^5\,e^{py} }{40 y^3}\,\left(1+\frac{3}{2py}+\frac{1}{p^2 y^2}\right)\,+\,\frac{p^6\,e^{-py}}{20 y^2}\,
 {\rm Ei}(2py)\,\left(1 + \frac{1}{py}\right)\,,
 \eea
 for the first few terms in the expansion ($p$ can be either positive or negative). Ei$(x)$ is the exponential integral function. The correct superposition of the Fourier-transformed solutions must reproduce ${\rm AdS}_5\times {\rm S}^5$ in the limit $Q_1\to 0$. This picks out the required combination,
 \be
 h_3(x,y)\,=\,\int dp\,\frac{e^{ipx}}{4}\,\left[\theta(p)\,p\, \tilde h_3(y;-p)\,-\,p\,\theta(-p)\, \tilde h_3(y;p)\right]\,.
 \ee 
In terms of the standard radial coordinate of AdS and the polar angle $\theta$ on the $S^5$ ($x=r\cos\theta$, $y=r\sin\theta$) to linear order in $Q_1$, we have, 
 \be
 h_3(r,\theta)\,=\,\frac{1}{r^4}\,\left(1+\,\frac{Q_1}{r^3}\,\frac{1-4\cos^2\theta}{4\sin^3\theta}\,+\,{\cal O}(Q_1^2)\right)\,.
 \ee
The dilaton, on the other hand, is exactly determined by 
\be
e^{\phi}\,=h_1^{-1/2}\,=\,\,\left(1+\frac{Q_1}{r^3\sin^3\theta}\right)^{-1/2}\,,
\ee
and vanishes at $\theta=0,\pi$ i.e. the north and south poles of the $S^5$ in the UV geometry, as expected for F1-string sources localized at $y=0$. It is instructive to look at the corrections to the metric components in the above expansion,
\be
g_{ii}\,=\,r^2 + \frac{Q_1}{r}\,\frac{1+4\cos^2\theta}{8\sin^3\theta}+{\cal O}(Q_1^2)\,,
\ee                                              
with similar corrections appearing for $g_{tt}$. The $1/r$ corrections to AdS asymptotics are characteristic of the backreaction due to string sources, as also noted in \cite{gundelman, headrick, Kumar:2012ui}. A similar analysis for the internal directions suggests that upon inclusion of the backreaction from the strings, the $S^4$ does not shrink at $\theta=0, \pi$, and that there are curvature singularities at these points.  We should stress, however, that we cannot trust the expansion in $Q_1$ in the vicinity of the string sources at $\theta=0,\pi$. Away from these points (fixed generic $\theta$ and large $r$) the geometry is asymptotically, locally ${\rm  AdS}_5\times {\rm S}^5$.

To summarize, we have argued in this section that the introduction of an ${\cal O}(N^2)$ density of heavy quarks (specifically, mutually BPS quark-antiquark pairs), preserving some supersymmetry in ${\cal N}=4$ SYM at large-$N$ and strong coupling, induces a flow from ${\rm AdS}_5\times {\rm S}^5$ to a scaling  Lifshitz-like IR background with $z=7$. The anisotropic scaling solution in the IR is somewhat novel due to the presence of {\em two} non-compact holographic directions, which scale differently to ensure that the resulting metric is scale invariant. 

\section{Coulomb branch solutions and hyperscaling violation}

A notable feature of the supersymmetric intersections discussed above is that they inherit the Coulomb branch moduli space of the ${\cal N}=4$ theory (where $SU(N)$ is Higgsed to $U(1)^{N-1}$ generically). In particular a probe D3-brane (spanning $t,x^1,x^2,x^3$) placed at any point in the $x$-$y$ plane, in the general background \eqref{ansatz}, experiences no force due to an exact cancellation between the Dirac-Born-Infeld (DBI) and Wess-Zumino terms in the probe action. A similar cancellation occurs (between the Nambu-Goto action and the coupling to $B_2$) for a probe F-string oriented parallel to the 
$x$-axis at any value of the $y$-coordinate. This points towards the existence of more general smeared solutions where both D3-branes and the F-strings are smeared with some distribution functions on the $x$-$y$ plane. We show that for  certain choices of such distribution functions the backreacted (IR) geometries can exhibit a range of Lifshitz-like scalings with hyperscaling violation.

General solutions with D3-branes on the Coulomb branch (within our $SO(5)$-symmetric ansatz) can be obtained by considering a general D3-brane density, so that
\be
h_3(x,y)\,=\,\int\int dx'\, dy' \frac{\rho_{\rm D3}(x',y')}{\left[(x-x')^2+(y-y')^2\right]^2}\,.
\ee
A general density function $\rho_{\rm D3}(x,y)$ preserves all supersymmetries whilst also leading to a violation of the Bianchi identity for $F_5$ due to the extended source distribution.

Since we want to interpret our solutions below as IR limits of Coulomb branch distributions it is useful to illustrate this with a simple example which allows to make contact with the F1-D3 intersection of \cite{Singh:2012un, deyroy}. Let us first consider a uniform distribution of D3-branes, in an interval along the $x$-axis with
\be
\rho_{\rm D3}(x,y)\,=\, \rho_{0}\,\delta(y)\,,\qquad\qquad 0\leq|x| \leq \tfrac{1}{2}\rho_0\,,
\ee
and no macroscopic string sources.
For large $(x,y)$ the geometry asymptotes to 
${\rm AdS}_5 \times
 {\rm S}^5$, whilst in the IR, for small $y$ and $|x| < \tfrac{1}{2}\rho_0$, we obtain
\be
h_3\,\simeq\,\frac{\pi}{2 y^3}\,\rho_0\,+\ldots.
\ee
This is a scaling regime where all metric components in \eqref{ansatz} only depend on powers of $y$.
Upon performing a reduction of \eqref{ansatz} on $S^4$ and the $x$-coordinate, the resulting 5D geometry precisely matches the reduction on a torus of the ${\rm AdS}_7 \times {\rm S}^4$ solution in 11D SUGRA \cite{HoyosBadajoz:2010td}. This $SO(5)$ symmetric Coulomb branch configuration can therefore be viewed as a flow at strong coupling and large-$N$, to an IR theory which appears to be (a subsector of) the 
6D superconformal theory with $(2,0)$ supersymmetry realized on a stack of M5-branes (compactified on a two-torus). The Coulomb branch configuration therefore deconstructs a higher dimensional field theory. In the deconstruction picture, the sizes of the extra dimensions are controlled by the inverse spacings between individual D3-branes on the Coulomb branch. The spacings $\sim {\cal O}(1/N)$ determine the masses of the lightest W-bosons and dyonic states on the Coulomb branch, which in turn are related to the Kaluza-Klein harmonics of the deconstructed compact dimensions. We would now like to understand how the IR dynamics on the Coulomb branch is modified by smeared, macroscopic  fundamental strings  viewed as heavy quarks in the gauge theory.

\subsection{Smeared F1-D3 intersections}

 In the presence of macroscopic string sources, both $h_1$ and $h_3$ will be non-trivial and the Bianchi identity for $F_5$ modified by D3-brane sources becomes
\be
\partial_y\left(y^4\partial_y h_3\right)\, +\,y^4\,h_1 \partial_x^2 h_3\,=\,-\rho_{\rm D3}(x,y)\,.
\ee
We will focus attention on distributions that, at least in 
some limit as in the above example, only depend on $y$, so that $\rho_{\rm D3}=\rho_{\rm D3}(y)$. Hence, for simplicity, we also assume that $h_3$ is a function of $y$ alone. 
Similarly the equation of motion for $H_3$ with a general smeared distribution of F-strings leads to the equation
\be
\frac{d}{dy}\left(y^4\frac{d h_1}{dy}\right)\,=\,-\rho_{\rm F1}(y).
\ee
As pointed out above, probe F-strings experience no force when they are aligned along the $x$-axis and placed at any value of $y$. Therefore a general $y$-dependent distribution should preserve
$1/4$-supersymmetry. In the Appendix we show that such smearing of sources satisfies a calibration condition which ensures that the string/brane embeddings we study respect  
the supersymmetries. The equations of motion follow from the type IIB supergravity action coupled to the smeared  D3-brane and string sources, 
\begin{equation}
S\,=\,S_{\rm IIB}+S_{\rm NG}+S_{\rm DBI} + \frac{1}{2\pi\alpha'}
\int B_2\wedge \Omega_8 \,+ \,4\, T_{\rm D3}\int C_4\wedge \Omega_6 .
\end{equation}
The actions for the sources are the smeared versions of the Nambu-Goto and DBI actions  with a particular choice of smearing forms $\Omega_8$ and $\Omega_6$, respectively
\bea
&&\Omega_8\,=\,-\frac{\pi\alpha^\prime}{\kappa^2}\,\rho_{\rm F1}(y)\,\,dy\wedge dx_1\wedge dx_2\wedge dx_3\wedge\epsilon_{(4)}\\\nonumber\\
&&\Omega_6\,=\,-(2\kappa^2\,T_{\rm D3})^{-1}\,\,\rho_{\rm D3}(y)\,\,dx\wedge dy\wedge\epsilon_{(4)}\,,
\eea
where we have defined $2\kappa^2\,\equiv 16\pi G_N$.
The sources alter the equations of motion for the dilaton, $H_3$, $F_5$ and the metric. In particular they contribute to the stress tensor. Imposing the weak energy condition (WEC) on the source stress tensor leads to positivity of the source density functions
\be
T_{AB}u^A u^B \geq 0\,,\qquad\implies \qquad {\rho}_{\rm D3}\,,
\rho_{\rm F1}\geq 0\,,
\ee
where $u_A$ is some timelike vector.
For more details on the equations of motion and their consistency with supersymmetry, we refer the reader to the Appendix.

To obtain scaling solutions, we simply choose power laws for the smearing densities
\begin{equation}
\rho_{\rm  F1}\,=\,\alpha\left(3-\alpha\right)\,Q_{1}\,y^{2-\alpha}\,,\qquad\qquad\qquad\rho_{\rm D3}\,=\,\beta\left(3-\beta\right)\,Q_{3}\,y^{2-\beta}\,.
\end{equation}
Then the equations of motion for $h_1$ and $h_3$ are also solved by power laws
\begin{equation}
h_1\,=\,\frac{Q_{1}}{y^\alpha}\,,\qquad\qquad\qquad h_3\,=\,\frac{Q_{3}}{y^\beta}\,.
\label{scale13}
\end{equation}
Positivity of the source densities requires $0\leq \alpha,\beta\leq 3\,$. Further imposing the null energy condition (NEC) on the complete stress tensor  leads to a weaker condition on the parameters $\alpha, \beta$ which is consistent with the WEC . The case $\alpha=\beta=3$ gives us the homogeneous solution of  \cite{Singh:2012un, deyroy}. This is the situation wherein the D3-branes and strings are all at $y=0$, but the branes are uniformly smeared along the $x$-axis.

 Inserting the general scaling solution into the metric (\ref{ansatz}) and reducing carefully over the $S^4$ and the $x$-coordinate (which we treat as a deconstructed compact dimension) to five-dimensional Einstein frame we obtain a family of Lifshitz geometries with hyperscaling violation:
\begin{equation}
ds_5^2\,=\,R^2\,r^{-\frac23\theta}\,\left(-r^{2z}dt^2+\frac{dr^2}{r^2}+r^2\,\sum_{i=1}^{3}dx^idx^i\right)\,.
\end{equation}
Here the radius is $R^2=\frac{4}{(\beta-2)^2}Q_{3}^{\frac12}Q_{1}^{\frac14}$ and we have changed the radial coordinate according to $y=r^{\frac{2}{\beta-2}}$ and rescaled the rest. The coefficients are related to the exponents of the sources as
\begin{equation}
z\,=\,1+\frac{\alpha}{\beta-2}\,,\qquad\qquad\qquad \theta\,=\,\frac{\alpha+4\beta-14}{\beta-2}
\end{equation}
 
\begin{figure}
\begin{center}
\includegraphics[width=3in]{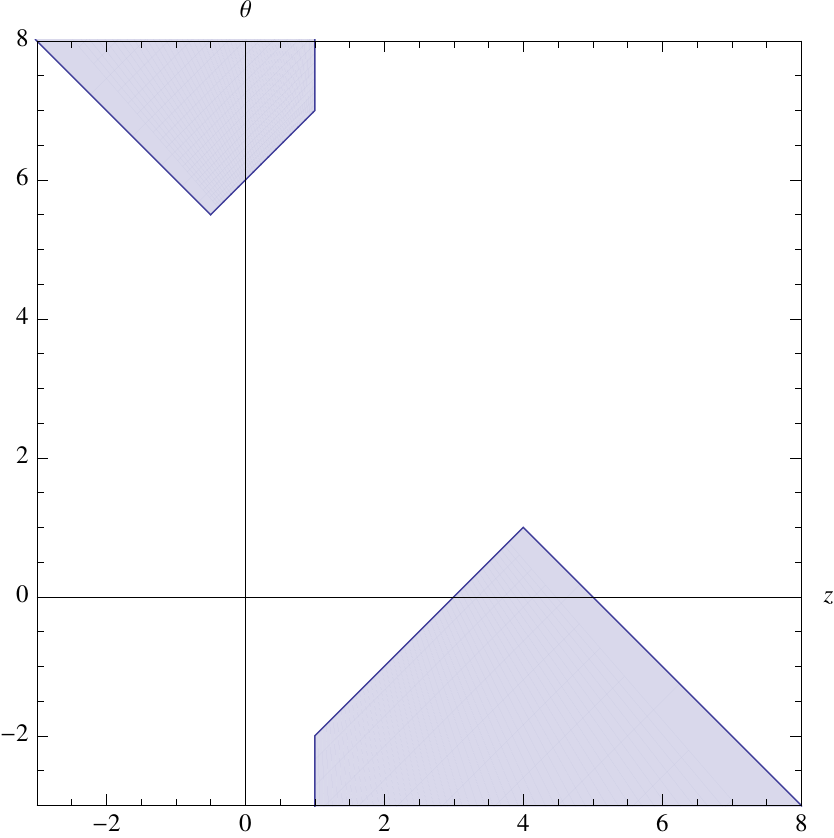}
\end{center}
\caption{\small {\label{zetatheta} The shaded region represents the allowed values for the dynamical exponent $z$ and hyperscaling violation coefficient $\theta$ for the solutions discussed in the text. If we impose an additional requirement that $\theta \leq d$ for stability (e.g. \cite{Dong:2012se}), this would also exclude the  shaded region top-left corner (which has $\beta<2$).}
}
\end{figure}

We stress that the emergence of hyperscaling violation along with Lifshitz scaling in the 4D spacetime is already evident in the ten dimensional metric \eqref{ansatz}, upon direct substitution of the scaling solutions \eqref{scale13}. Reduction to 5D changes the hyperscaling violation parameter, but leaves the Lifshitz exponent unchanged. 

The case $\beta=2$  needs to be treated separately, as we do below. Given that the weak energy conditions for the sources, considered separately, restrict the values of the parameters to lie in the range $0\le\alpha,\,\beta\le3$, the allowed values for $z$ and $\theta$ in this class of solutions are shown in Fig. \ref{zetatheta}.  Note that if we also require $\theta \leq d$ for stability as argued in \cite{Dong:2012se}, only the solutions shown in the lower right corner of Fig. \ref{zetatheta} with 
$\theta \leq 1$ survive. These correspond to D3-brane distributions with $\beta >2$.  Finally, there are no solutions with $\theta=2$, which would lead to a logarithmic violation of the area law for entanglement entropy.

\paragraph{\underline{$\beta=2$: Conformally ${\rm AdS}_2\times {\mathbb R}^3$}:}As stated before, the case $\beta=2$ has to be treated separately, since the change of variables used for the radial coordinate is not well defined for that value of the parameter. Importantly as $\beta \to 2$, both $z$ and $\theta$
diverge, 
\be
\lim_{\beta \to 2}\quad z,\theta\to\infty\,
\qquad \qquad\lim_{\beta \to 2}\quad\eta\equiv -\tfrac{\theta}{z}\to \left(\tfrac{6}{\alpha}-1\right)\,.
\ee
In the allowed range for $\alpha$, we have $\eta \geq 1$.
The $\beta\to 2$ limit is also interesting because it corresponds to a uniform distribution of D3-branes on the $x$-$y$ plane,
\begin{equation}
\rho_{\rm D3}\,=\,2\,Q_{3}\,,
\end{equation}
so that the harmonic function is,
\begin{equation}
h_3\,=\,\frac{Q_{3}}{y^2}\,.
\end{equation}
and $h_1=Q_1/y^\alpha$ as before. Hence we  obtain a 
one-parameter ($\alpha\ne0$) family of solutions. Again, substituting this into the ten-dimensional metric and reducing to the five-dimensional Einstein frame with the change of variable  $y=r^{\frac{2}{\alpha}}$ we  arrive at a metric conformal to AdS$_2\times\mathbb{R}^3$,
\begin{equation}
ds_5^2\,=\,R^2\,r^{\frac23 \eta}\,\left(-r^2dt^2+\frac{dr^2}{r^2}+\sum_{i=1}^{3}dx^idx^i\right)\,,
\end{equation}
where the radius is now $R^2=4\alpha^{-2}Q_{3}^{\frac12}Q_{1}^{\frac14}$. Geometries with an IR factor conformal to 
${\rm AdS}_2\times {\mathbb R}^2$ have been used to describe locally quantum critical theories and argued to have certain properties desirable for a holographic description of Fermi surfaces \cite{Hartnoll:2012wm}. The zero temperature entropy density associated to the conformally ${\rm AdS}_2\times {\mathbb R}^3$ background is vanishing since $\eta>0$ and the entanglement entropy has no extensive finite contribution. As pointed out in 
\cite{Hartnoll:2012wm}, the entanglement entropy of a strip in this case has a rather puzzling behaviour in that the extremal surface only exists for a very specific value of the strip width.

In all the geometries, including the conformally ${\rm AdS}_2\times {\mathbb R}^3$ backgrounds, curvature singularities appear in the deep IR (see e.g.\cite{Horowitz:2011gh}). In particular, since the dilaton is given by $e^\phi\,=\, h_1^{-1/2}$, and runs to zero in the IR, the string frame metric leads to  divergent curvatures  and the solutions will be expected to receive large $\alpha^\prime$ corrections. One can S-dualize the backgrounds, and then the problem becomes one of strong coupling in the IR.
It is a compelling question to ask what low energy physics arises and resolves such curvature singularities
\cite{Harrison:2012vy, Bhattacharya:2012zu}. This may be an interesting avenue to explore since we have some knowledge of the microscopic field theory the geometries describe.

\section{Discussion}

The supersymmetric scaling geometries we have discussed in this paper were obtained from previously known intersecting brane configurations along with a new ingredient, namely, generic source distributions compatible with the supersymmetries.  Our motivation was to understand whether any universal features emerge within a tractable holographic framework, when a state of finite (quark)
 density is introduced into a known field theory (with a large-$N$ string/gravity dual). The physical interpretation of the setup explored in this paper is that it corresponds to a uniform density of mutually BPS quarks and anti-quarks (with opposite $SO(6)$ orientations) in ${\cal N}=4$ SYM. In particular, these configurations all have vanishing three form flux, $F_3$. The presence of  non-zero $F_3$ can be interpreted as a finite baryon density, as was the case in \cite{Kumar:2012ui} (see also
\cite{Chen:2009kx} for related discussions). In a forthcoming publication \cite{paper2} we will derive the general form for the $\tfrac{1}{4}$-BPS configurations involving smeared strings with non-vanishing $F_3$.

One of the larger aims of this study is to understand whether the simplistic picture of backreacting smeared heavy quarks can be embedded into holographic setups with dynamical flavours, most notably the smeared D3-D7 system explored, for instance, in \cite{Bigazzi:2011it, Bigazzi:2013jqa, CarlosReview}.

An important observation in this paper is  the emergence of a $z=7$ IR scaling regime from the partially localized intersection described by homogeneous (sourceless) equations. Despite having different global symmetries and no supersymmetry, the same scaling was observed in \cite{Kumar:2012ui, Azeyanagi:2009pr}. The value of $z$ can now be understood within the intersecting brane framework. Following the results of \cite{Youm:1999zs}, it is easy to verify that, more generally, partially localized F1-Dp intersections give rise to anisotropic scaling with  $z=\frac{16-3p}{4-p}$, when $p< 4$.

Finding finite temperature, black brane generalizations of the solutions discussed in this paper appears difficult. For the partially localized intersection  (with $z=7$), the solutions are functions of {\em two} coordinates, $v = x^2y$ and $y$, and this makes the problem of `blackening' the solutions challenging. We leave this question for future investigation. When the sources are smeared, as for the Coulomb branch configurations, the zero temperature delocalization is possible only because the theory has a moduli space of vacua. Finite temperature lifts such moduli spaces so that the free energy is minimized at the origin of moduli space\footnote{Away from the origin, various degrees of freedom are rendered massive, and  their contribution to the entropy thus decreases.}. A possible approach towards the finite temperature physics of smeared Coulomb branch distributions is to introduce chemical potential for charges under the global symmetry group, i.e. to consider rotating brane configurations along the lines of \cite{Kraus:1998hv, Brandhuber:1999jr}.

Within the context of the Coulomb branch intersctions, it is interesting to note that by engineering appropriate string/D3-brane sources we should be able to obtain zero temperature flows between different scaling geometries.

\paragraph{Acknowledgments:} We would like to thank Paolo Benincasa, Jerome Gaillard, Tim Hollowood, David Mateos, Carlos Nu$\tilde{\rm n}$ez, Andy O'Bannon, Alfonso Ramallo and Javier Tarrio for illuminating discussions and 
helpful comments on various aspects of this work. We acknowledge partial support from the STFC grant ST/G000506/1. The work of A. F. was supported by STFC grant ST/J00040X/1.

\newpage

\startappendix
\Appendix{Calibration conditions and equations of motion}
We work in Einstein frame in the conventions of \cite{D'Hoker:2007fq}. We restrict our discussion to the situation with $F_3=0$, and $y$-dependent density distributions and harmonic functions. Smearing the fundamental strings along all the transverse directions, the Nambu--Goto action plus coupling to the NS form are schematically,
\begin{equation}
S_{\rm F1}\,=\,-\,\int\,\left(e^{\frac{\phi}{2}}\,\mathcal{K}_2-B_2\right)\wedge\Omega_8\,,
\end{equation}
with a particular choice of smearing form 
\begin{equation}
\Omega_8\,=\,-\rho_{\rm F1}(y)\,\,dy\wedge dx_1\wedge dx_2\wedge dx_3\wedge\epsilon_{(4)}\,.
\end{equation}
The function of the radial coordinate $\rho_{\rm F1}(y)$ describes the string charge distribution. Associated to a string world-sheet embedding, one  introduces a calibration form, given essentially by the induced metric on the string, 
\begin{equation}
\mathcal{K}_2\,=\,-h_1^{-\frac34}\,dt\wedge dx\,.
\end{equation}
In the presence of smeared sources the calibration condition, which ensures that the embedding of the branes/strings respects supersymmetry, has to be modified (see for instance \cite{CarlosReview}). In the case of fundamental strings  the modified condition is
\begin{equation}
d\left(e^{\frac{\phi}{2}}\,\mathcal{K}_2\right)\,=\,H_3\,.
\end{equation}
Using that $h_1=e^{-2\phi}$, one can easily check that this condition is verified in our ansatz, establishing that  general backreacting string distributions are consistent with the supersymmetries of the setup. 
Similarly, we introduce a set of D3 branes extending along the 4D Minkowski directions and smeared on the transverse coordinates 
, with action 
\begin{equation}
S_{\rm D3}\,=\,-\,\int\,\left(\mathcal{K}_4-4C_4\right)\wedge\Omega_6\,.
\end{equation}
In this case the smearing form is
\begin{equation}
\Omega_6\,=\,-\rho_{\rm D3}(y)\,\,dx\wedge dy\wedge\epsilon_{(4)}\,,
\end{equation}
while the calibration form reads
\begin{equation}
\mathcal{K}_4\,=\,-h_3^{-1}\,dt\wedge dx_1\wedge dx_2\wedge dx_3\,,
\end{equation}
where again the function $\rho_{\rm D3}(y)$ parametrizes the brane charge along the radial direction. The pertinent calibration condition
\begin{equation}
d\mathcal{K}_4\,=\, 4dC_4
\end{equation}
is also straightforwardly satisfied, again in line with the expected supersymmetry of the backgrounds.  

The presence of these smeared sources alters several equations of motion, that now read
\begin{eqnarray}
&& d\left(e^{-\phi}*H_3\right)+\Omega_8=0\nonumber\\\nonumber\\
&& d*F_5+\tfrac{1}{4}\Omega_6=0\nonumber\\\nonumber\\
&& d*d\phi+\frac12e^{-\phi}\,H_3\wedge*H_3-\frac12e^{\frac{\phi}{2}}\,\mathcal{K}_2\wedge\Omega_8=0\nonumber\\\nonumber\\
&&G_{MN}=T_{MN}^{\rm IIB}+T_{MN}^{\rm F1}+T_{MN}^{\rm D3}
\end{eqnarray}
The equations for $F_1$ and $F_3$ are not modified and turn out to be automatically satisfied within our ansatz. Notice that $B_2$ couples electrically to the strings through the Nambu--Goto action, so the Bianchi $dH_3=0$ remains intact. As is customary the Bianchi for $F_5$ coincides with its equation of motion. The stress tensors for the smeared sources are
\begin{eqnarray}
T_{MN}^{\rm F1}&=&-\frac{1}{\sqrt{-g}}\frac{\delta S_{\rm F1}}{\delta g^{MN}}\,=\,\frac12\,e^{\frac{\phi}{2}}\Big(\,g_{MN}\,\Omega_8\lrcorner\left(*\mathcal{K}_2\right)-\iota_{(M}\Omega_8\lrcorner\iota_{N)}\left(*\mathcal{K}_2\right)\Big)\nonumber\\[2mm]
T_{MN}^{\rm D3}&=&-\frac{1}{\sqrt{-g}}\frac{\delta S_{\rm D3}}{\delta g^{MN}}\,=\,\frac12\,\Big(\,g_{MN}\,\Omega_6\lrcorner\left(*\mathcal{K}_4\right)-\iota_{(M}\Omega_6\lrcorner\iota_{N)}\left(*\mathcal{K}_4\right)\Big)
\end{eqnarray}
where for arbitrary $p$-forms $\omega_p$ and $\xi_p$ we have defined the operations
\begin{eqnarray}
\iota_M\omega_p&\equiv&\frac{1}{(p-1)!}(\omega_p)_{MN_1\cdots N_{p-1}}\,dx^{N_1}\wedge \cdots \wedge dx^{N_{p-1}}\nonumber\\
\omega_p \lrcorner \xi_p& \equiv &\frac{1}{p!}(\omega_p)_{M_1\cdots M_p} (\xi_p)^{M_1\cdots M_p}
\end{eqnarray}
Using the known ansatz for the forms, all the equations above boil down to the equations determining the harmonic functions sourced by the charge distributions
\begin{eqnarray}
\frac{d}{dy}\left(y^4\frac{dh_1}{dy}\right)&=&-\rho_{\rm F1}(y)\,,\nonumber\\[2mm]
\frac{d}{dy}\left(y^4\frac{dh_3}{dy}\right)&=&-\rho_{\rm D3}(y)\,.
\end{eqnarray}

\end{document}